\documentclass[aps,prd,showpacs,twocolumn,amsmath]{revtex4-1}
\usepackage[colorlinks,urlcolor=blue,linkcolor=blue,citecolor=blue]{hyperref} %
\usepackage{epsfig,graphicx,amsmath,amssymb,float}
\usepackage{color} 
\usepackage{ulem}  
\begin{document} 

\title{{\bf Will the subprocesses $\rho(770,1450)^0\to K^+K^-$ contribute large  branching \\ 
               fractions for $B^\pm \to \pi^\pm K^+K^-$ decays?} }
                    
\author{Wen-Fei Wang}\email{wfwang@sxu.edu.cn}   
\affiliation{\vspace{0.2cm} Institute of Theoretical Physics, Shanxi University, Taiyuan, Shanxi 030006, China \vspace{0.2cm}}  

\date{\today}

\begin{abstract}
We analyze the quasi-two-body decays $B^+\to\pi^+\rho(770,1450)^0 \to\pi^+ K^+K^-$ in the perturbative QCD approach. 
The results in this work do not support that large branching fractions contributed by the resonances $\rho(770,1450)^0$ 
in the $B^\pm \to \pi^\pm K^+K^-$ decays. The virtual contribution for $K^+K^-$ from the tail of the resonance $\rho(770)^0$ 
which has been ignored in the experimental studies is about $1.5$ times of the $\rho(1450)^0 \to K^+K^-$ contribution, with the
predicted branching fractions ${\mathcal B}_v=(1.31\pm0.27)\times10^{-7}$ and ${\mathcal B}=(8.96\pm2.61)\times10^{-8}$, 
respectively, for these two subprocesses in the $B^\pm \to \pi^\pm K^+K^-$ decays. The absence of $\rho(770)^0\to K^+K^-$ 
for the decay amplitude of a three-body hadronic $B$ decay involving charged kaon pair could probably result in a larger 
proportion for the contribution from the resonance $\rho(1450)^0$ in experimental analysis.
\end{abstract}

\maketitle

\section{Introduction}

Strong dynamics in the charmless three-body hadronic $B$ meson decays is related to the short-distance processes like in the 
two-body cases, and it also involves the hadron-hadron interactions, the three-body effects~\cite{npps199-341,prd84-094001} and 
the rescattering processes~\cite{1512-09284,prd89-094013} in the final states. In experimental studies, the strong interactions 
along with the weak processes of a three-body $B$ meson decay are always described in its decay amplitude as the coherent 
sum of the resonant and nonresonant contributions in the isobar formalism~\cite{pr135-B551,pr166-1731,prd11-3165}. 
The isobar expression with or without certain resonances will certainly have impacts on the fit fractions, and then influence 
the observables such as the branching ratios and $CP$ violations for the resonant and the nonresonant contributions 
of the three-body process in view of the explicit distribution of the experimental events in the Dalitz plot~\cite{dalitz}.

Very recently, LHCb Collaboration presented a surprising large fit fraction $(30.7\pm1.2\pm0.9)\%$ of the total three-body result 
in~\cite{prl123-231802} for the subprocess $\rho(1450)^0\to K^+K^-$ in the amplitude analysis of the $B^\pm\to \pi^\pm K^+K^-$ 
decays. Considering the branching fraction $(5.2\pm0.4)\times10^{-6}$ for $B^+\to K^+K^-\pi^+$ in the {\it Review of Particle 
Physics}~\cite{PDG-2018} which was averaged from the results $(5.38\pm0.40\pm0.35)\times10^{-6}$ in~\cite{prd96-031101} provided 
by Belle and $(5.0\pm0.5\pm0.5)\times10^{-6}$ in~\cite{prl99-221801} presented by BaBar, one has the branching fraction 
$(1.60\pm0.15)\times10^{-6}$ for the quasi-two-body decay $B^+\to\pi^+\rho(1450)^0 \to\pi^+ K^+K^-$ from LHCb's fit fraction.
While the contributions of $B^\pm\to\pi^\pm\rho(1450)^0$ were found quite small in the $\rho^0$ dominant decay modes 
$B^\pm\to \pi^\pm \pi^+\pi^-$ in~\cite{prl124-031801,prd101-012006} by LHCb recently.

The resonance contributions for the charged kaon pair in the three-body decays $B^\pm \to \pi^\pm K^+K^-$ are associated with the 
low energy scalar, vector and tensor states, such as $f_0(980)$, $\phi(1020)$, $f^\prime_2(1525)$, etc., which have been noticed 
in~\cite{prd76-094006,prd88-114014,2003-03754,epjc79-792}, while the contributions for $K^+K^-$ also come from the $P$-wave 
resonances $\rho(770)$, $\omega(782)$ and their excited states~\cite{prd96-113003,prd67-034012}. The natural mode for $
\rho(770)^0$ decays into $K^+K^-$ is blocked because of its pole mass which is below the threshold of kaon pair,  but the virtual 
contribution~\cite{prd94-072001,plb791-342} from the tail of the Breit-Wigner (BW) formula~\cite{BW-model} of $\rho(770)$ 
could be indispensable for $K^+K^-$ in the $B^\pm\to \pi^\pm K^+K^-$ decays, which could be deduced from the works for the 
processes of $\pi^-p\to K^-K^+n$ and $\pi^+n\to K^-K^+p$~\cite{prd15-3196,prd22-2595}, 
$e^+e^- \to K^+K^-$~\cite{plb669-217,prd88-032013,prd94-112006,plb779-64,prd99-032001} and $\pi \pi \to K\bar K$ 
scattering~\cite{epjc78-897}. 

\begin{figure*}[t]
  \begin{center}
    \includegraphics[width=0.8\textwidth]{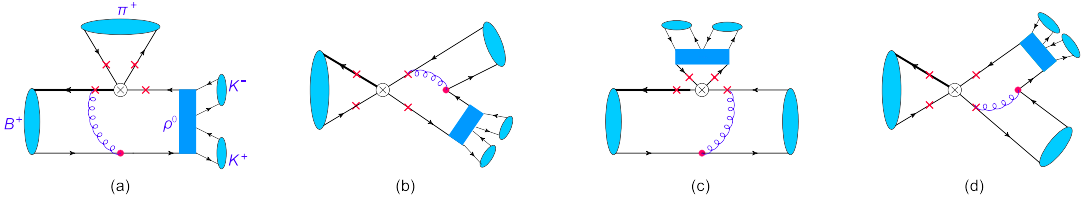}
  \end{center}
  \vspace{-0.2cm}
\caption{Feynman diagrams for the processes $B^+\to \pi^+ \rho^0 \to \pi^+ K^+ K^-$, with $\otimes$ is the 
               weak vertex, $\times$ denotes possible attachments for the hard gluons and the rectangle represents the 
               resonance $\rho(770)^0$ and its excited states.}
\label{fig-feyndiag}
\end{figure*}

We shall analyze the $\rho(770,1450)^0\to K^+K^-$ contributions for the three-body decays $B^\pm\to \pi^\pm K^+K^-$ with the 
quasi-two-body framework based on the perturbative QCD (PQCD) approach~\cite{plb504-6,prd63-054008,prd63-074009,ppnp51-85}. 
The final state interaction effect was found to be suppressed for the $\rho^0\to K^+K^-$ process~\cite{prd75-074017} and will be 
neglected in the calculation in this work. For the quasi-two-body decays $B^\pm\to\pi^\pm\rho^0\to \pi^\pm K^+K^-$, the 
intermediate states $\rho(770)^0$ and $\rho(1450)^0$ are generated in the hadronization of the light quark-antiquark pair $q\bar q$, 
with $q$=$(u,d)$, as demonstrated in the Fig.~\ref{fig-feyndiag}, in which the factorizable and nonfactorizable diagrams have been 
merged for the sake of simplicity and Fig.~\ref{fig-feyndiag}~(b) and~(d) will generate different decay amplitudes because of the 
different position of the bachelor and the intermediate states. The subprocesses $\rho(770,1450)^0\to K^+K^-$, which can not be 
calculated in the PQCD approach, could be introduced into the distribution amplitudes of the kaon pair system by the time-like form factor of kaon. The quasi-two-body framework based on PQCD approach has been discussed in detail in~\cite{plb763-29}, and has 
been adopted in some works for the quasi-two-body $B$ meson decays~\cite{jhep2003-162,plb788-468,prd96-036014,prd95-056008,
prd96-093011,npb923-54} recently. 

This paper is organized as follows: In Sec.~\ref{sec-2}, we briefly review the electromagnetic time-like form factor for the charged 
kaon, we introduce the $P$-wave $K^+K^-$ system distribution amplitudes and give the expression of differential branching fractions 
for the quasi-two-body decays $B^\pm\to\pi^\pm\rho^0\to \pi^\pm K^+K^-$. In Sec.~\ref{sec-3}, we provide numerical results for the 
concerned decay processes and give some necessary discussions. The conclusions are presented in Sec.~\ref{sec-sum}.

\section{Framework}\label{sec:2}
\label{sec-2}
In the light-cone coordinates, the momentum $p_B$ and the light spectator quark's momentum $k_B$ for $B^+$ are defined as 
\begin{eqnarray}
p_B=\frac{m_B}{\sqrt2}(1,1,0_{\rm T}),\quad  k_B=\left(\frac{m_B}{\sqrt2}x_B, 0, k_{B{\rm T}}\right),
\end{eqnarray} 
in the rest frame of $B$ meson, where $m_B$ is the mass of $B^+$.
The resonance $\rho$ has the momentum $p=\frac{m_B}{\sqrt 2}(\eta, 1, 0)$,  with $\eta=s/m_B^2$ and $s=p^2$,  
its spectator quark got the momentum $k=(0, \frac{m_B}{\sqrt 2}z, k_{\rm T})$. 
For the bachelor final state pion and its spectator quark, the momenta $p_3$ and $k_3$ have the definitions 
\begin{eqnarray}
p_3=\frac{m_B}{\sqrt2}(1-\eta, 0, 0_{\rm T}),\;
k_3=\left(\frac{m_B}{\sqrt2}(1-\eta)x_3, 0, k_{3{\rm T}}\right)\!.\quad
\end{eqnarray}
The $x_B$,  $x_3$ and $z$ are the momentum fractions which run from zero to one in the calculation.

The time-like form factor $F_{K}$ for the charged kaon is related to the electromagnetic form factor and defined by~\cite{epjc39-41}
\begin{eqnarray} 
  \langle K^+(p_1) K^-(p_2) | j^{em}_\mu | 0 \rangle = (p_1-p_2)_\mu \,F_{K}(s), 
\end{eqnarray}
with the squared invariant mass $s=(p_1+p_2)^2$ for kaon pair and the constraint $F_{K}(0)=1$. The electromagnetic current 
$j^{em}_\mu=\frac23\bar u\gamma_\mu u-\frac13\bar d\gamma_\mu d -\frac13\bar s\gamma_\mu s$ is carried by the light 
quarks~\cite{npb250-517}. With the BW formula for the resonances $\omega$ and $\phi$ and the Gounaris-Sakurai (GS) 
model~\cite{prl21-244} for $\rho$, we have the time-like form factor~\cite{epjc39-41}
\begin{eqnarray} 
  F_{K}(s)&=&\frac12\sum_\rho c^K_\rho {\rm GS}_\rho(s) +\frac16\sum_\omega c^K_\omega {\rm BW}_\omega(s) \nonumber\\
               &+&\frac13\sum_\phi c^K_\phi {\rm BW}_\phi(s),  
  \label{def-F-K+}
\end{eqnarray}
where the $\sum$ means the summation of the resonances $\rho, \omega$ or $\phi$ and their corresponding excited states, the 
explicit expressions and auxiliary functions for BW and GS are referred to Refs.~\cite{prd86-032013,prl21-244}.  
The parameters $c^K_{\rho(\omega,\phi)}$ have been fitted to the data in Refs~\cite{epjc39-41,prd81-094014,jetp129-386}.

For the concerned subprocesses $\rho(770,1450)^0\to K^+K^-$, the $P$-wave $K^+K^-$ system distribution amplitudes are 
organized into~\cite{prd76-074018} 
\begin{eqnarray} 
   \phi^{P\text{-wave}}_{KK}(z,s)&=&\frac{-1}{\sqrt{2N_c}}\big[\sqrt{s}\,{\epsilon\hspace{-1.5truemm}/}_L\phi^0(z,s) +  
                                                             {\epsilon\hspace{-1.5truemm}/}_L {p\hspace{-1.7truemm}/} \phi^t(z,s)   \nonumber\\
                                                   &+& \sqrt s \phi^s(z,s)  \big].
\end{eqnarray}
The momentum $p=p_1+p_2$, and $\epsilon_L$ is the longitudinal polarization vector for the resonances.  
We have the distribution amplitudes as~\cite{plb763-29}
\begin{eqnarray}
   \phi^{0}(z,s)&=&\frac{3F^\rho_K(s)}{\sqrt{N_c}} z(1-z)\left[1+a_2^{0} C^{3/2}_2(1-2z) \right]\!,\label{def-DA-0}\\
   \phi^{t}(z,s)&=&\frac{3F^t_K(s)}{2\sqrt{N_c}}(1-2z)^2\left[1+a_2^t  C^{3/2}_2(1-2z)\right]\!,\label{def-DA-t}\\
   \phi^{s}(z,s)&=&\frac{3F^s_K(s)}{2\sqrt{N_c}}(1-2z)\left[1+a_2^s\left(1-10z+10z^2\right) \right]\!,\label{def-DA-s}\quad
\end{eqnarray}
where the $F^\rho_K$ is the $\rho$ term of Eq.~(\ref{def-F-K+}), the Gegenbauer polynomial 
$C^{3/2}_2(\chi)=3\left(5\chi^2-1\right)/2$. The Gegenbauer moments $a_2^{0}=0.25\pm0.10$, $a_2^t=-0.60\pm0.20$ and 
$a_2^s=0.75\pm0.25$ which are the same as they in~\cite{plb763-29} for the intermediate state $\rho$. For the subprocess 
$\rho^0\to K^+K^-$, we adopt $F^{s,t}_K(s)\approx (f^T_{\rho}/f_{\rho})F^\rho_K(s)$ as in~\cite{plb763-29} for 
the channel $\rho^0\to \pi^+\pi^-$.

One has the differential branching fractions ($\mathcal B$) for the quasi-two-body decays 
$B^\pm\to\pi^\pm\rho^0\to \pi^\pm K^+K^-$ as~\cite{prd74-114009,prd79-094005,plb791-342}
\begin{eqnarray}
  \frac{d{\mathcal B}}{d\eta}=\tau_B\frac{|\overrightarrow{q_\pi}|^3 |\overrightarrow{q_K}|^3}{12\pi^3m^5_B}\overline{|{\mathcal A}|^2},
\label{eqn-def-br}
\end{eqnarray}
where $\tau_B$ being the $B$ meson mean lifetime. It should be noted that the Eqs.~(\ref{def-DA-0})-(\ref{def-DA-s}) and 
Eq.~(\ref{eqn-def-br}) are slightly different from the corresponding expressions in~\cite{plb763-29}. These differences are caused 
by the introduction of the Zemach tensor ${\small -2\overrightarrow{q_\pi}}\cdot{\small \overrightarrow{q_K}}$~\cite{Zemach} in 
this work as did in Refs.~\cite{prd74-114009,prd79-094005}, this tensor is employed to describe the angular distribution for the decay 
of spin $1$ resonances.  The magnitudes of the kaon and pion 
momenta $|\overrightarrow{q_K}|$ and $|\overrightarrow{q_\pi}|$ are written, in the center-of-mass frame of the kaon pair, as
\begin{eqnarray}
 |\overrightarrow{q_K}|  &=&\frac12\sqrt{s-4m^2_{K}}, \label{def-q-K} \\
 |\overrightarrow{q_\pi}|&=&\frac{1}{2\sqrt s}\sqrt{\left(m^2_{B}-m_{\pi}^2\right)^2 -2\left(m^2_{B}+m_{\pi}^2\right)s+s^2}, \quad
 \label{def-q-pi}
\end{eqnarray}
with the pion mass $m_\pi$ and the kaon mass $m_K$.     

The decay amplitudes ${\cal A}$ for the quasi-two-body decays $B^+\to\pi^+\rho(770,1450)^0\to \pi^+ K^+K^-$ 
dependent only on the quark structures of the hadronic matrix elements for $B^+$ to $\pi^+\rho(770,1450)^0$ transitions and 
have the same expressions as the decays $B^+\to\pi^+\rho(770,1450)^0\to \pi^+ \pi^+\pi^-$ in~\cite{prd96-036014,prd95-056008}
except the replacement of the form factor $F_\pi\to F_K$. The amplitudes in ${\cal A}$ according to the diagrams in 
Fig.~\ref{fig-feyndiag} could be found in the Appendix of~\cite{plb763-29}.

\section{Results}\label{sec:3}
\label{sec-3}
In the numerical calculation, we adopt the decay constant $f_B=0.189$ GeV~\cite{prd98-074512} and 
the mean lifetime $\tau=(1.638\pm0.004)\times 10^{-12}$~s~\cite{PDG-2018} for the $B^+$ meson. 
The masses and the decay constant $f_\pi$ for the relevant particles in the numerical calculation in this work, the full widths for 
$\rho(770)$ and $\rho(1450)$, and the Wolfenstein parameters of the CKM matrix are presented in Table~\ref{tab1}.

\begin{table}[H]  
\begin{center}
\caption{Masses, decay constant $f_\pi$, full widths of $\rho(770)$ and $\rho(1450)$ (in units of GeV) and Wolfenstein 
parameters~\cite{PDG-2018}.}
\label{tab1}
\begin{tabular}{l}\hline\hline 
\;$m_{B^{\pm}}=5.279 \quad\;\,\; m_{\pi^\pm}=0.140 \quad m_{K^\pm}=0.494  $ \\  
\;$m_{\rho(770)}=0.775 \;\;\;  m_{\rho(1450)}=1.465 \;\;\;   f_\pi=0.131 $ \\ 
\;$\Gamma_{\rho(770)}=0.149\;\quad \Gamma_{\rho(1450)}=0.400\pm0.060  $   \\ 
\;$\lambda=0.22453\pm 0.00044  \quad\;\;\;  A=0.836\pm0.015\; $ \\
\;$\bar{\rho} = 0.122^{+0.018}_{-0.017} \hspace{2.0cm} \bar{\eta}= 0.355^{+0.012}_{-0.011} $\\ 
\hline\hline  
\end{tabular}
\end{center}
\end{table}

The ratio between the $f^T_\rho$ and $f_\rho$ for $\rho(770)$ has been computed in lattice QCD, we choose the result 
$f^T_\rho/f_\rho=0.687$ at the scale $\mu=2$ GeV~\cite{prd78-114509} for our numerical calculation. The decay constant 
$f_\rho$ for $\rho(770)$ can be extracted from the processes $\tau^-\to\rho^-\nu_\tau$ and $\rho^0\to e^+e^-$, we take the 
value $0.216$ GeV~\cite{jhep1608-098}. The $c^K_{\rho(770)}$ was fitted to be $1.139\pm0.010$ and $1.195\pm0.009$ with the 
unconstrained and constrained fit procedure in~\cite{epjc39-41}, respectively, which are consistent with the values 
$1.138\pm0.011$ and $1.120\pm0.007$ in~\cite{prd81-094014}. We employ the result $1.139\pm0.010$ for the quasi-two-body 
decay $B^+\to\pi^+\rho(770)^0\to \pi^+ K^+K^-$. As for the $c^K$ for $\rho(1450)$, we adopt the value 
$-0.124\pm0.012$~\cite{epjc39-41} for the subprocess $\rho(1450)^0\to K^+K^-$, which is close to the constrained fit result$-0.112\pm0.010$ in~\cite{epjc39-41} and the unconstrained fit result $-0.107\pm0.010$ in~\cite{prd81-094014}.
We need to stress that different choices for $c^K_{\rho(770)}$ and $c^K_{\rho(1450)}$ with the values in~\cite{epjc39-41,
prd81-094014} can not significantly change the numerical results and will not change our conclusions in this work.

Utilizing the differential branching fractions the Eq.~(\ref{eqn-def-br}), we have the branching fractions
\begin{eqnarray}   
   {\mathcal B} (B^\pm&\to&\pi^\pm\rho(1450)^0\to \pi^\pm K^+K^-)                               \nonumber \\
                                    &=&\big(  8.96\pm1.58(a^s_2+a^0_2+a^t_2) \pm0.83(\omega_B)  \nonumber\\  
                                    &\pm& 0.79(m^\pi_0+a^\pi_2)  \pm1.73(c^K_{\rho(1450)}) \big) \times10^{-8}, \label{res-br-pi+1450}\quad \\
  {\mathcal B}_v(B^\pm&\to&\pi^\pm\rho(770)^0\to \pi^\pm K^+K^-)                                \nonumber\\  
                                   &=& \big(1.31\pm0.22(a^s_2+a^0_2+a^t_2) \pm0.12(\omega_B)     \nonumber\\  
                                   &\pm& 0.11(m^\pi_0+a^\pi_2)   \pm0.02(c^K_{\rho(770)})  \big) \times10^{-7}.  \label{res-br-pi+R770}
\end{eqnarray}   
The subscript $v$ for ${\mathcal B}$ of $B^\pm\to\pi^\pm\rho(770)^0\to \pi^\pm K^+K^-$ means the virtual 
contribution~\cite{prd94-072001,plb791-342} which is integrated of the Eq.~(\ref{eqn-def-br}) from the threshold of kaon pair. 
The first error for the two branching fractions above comes from the uncertainty of the Gegenbauer moments $a_2^s, a_2^{0}$ 
and $a_2^t$ in the $K^+K^-$ system distribution amplitudes, the second error is induced by the shape parameter 
$\omega_B=0.40\pm0.04$ for $B^+$, the third one is contributed by the chiral mass $m^\pi_0=1.40\pm0.10$ GeV and the 
Gegenbauer moment $a^\pi_2=0.25\pm0.15$ for pion and the fourth one due to the variation of $c^K$ in the form factor $F_K$.
There are other errors come from the uncertainties of the Wolfenstein parameters of the CKM matrix, the parameters 
in the distribution amplitudes of bachelor pion, the masses and the decay constants of the initial and final states, etc. 
are small and have been neglected.

From the fit fractions in~\cite{prl123-231802}, one has a branching fraction ${\mathcal B}\approx(1.60\pm0.15)\times10^{-6}$ for the 
quasi-two-body decay $B^+\to\pi^+\rho(1450)^0 \to\pi^+ K^+K^-$, which is about $18$ times larger than the PQCD prediction 
shown in Eq.~(\ref{res-br-pi+1450}). The value $(1.60\pm0.15)\times10^{-6}$ is close to the result 
${\mathcal B}=1.4^{+0.6}_{-0.9}\times10^{-6}$ for the decay $B^+\to\rho(1450)^0\pi^+$ with $\rho(1450)^0\to\pi^+\pi^-$ from 
BaBar Collaboration~\cite{prd79-072006,PDG-2018}. Recently, in the amplitude analysis of the $B^+\to\pi^+\pi^+\pi^-$ decay, 
LHCb Collaboration provided a $CP$-averaged fit fraction $(5.2\pm0.3\pm0.2\pm1.9)\%$~\cite {prl124-031801,prd101-012006} for 
the $\rho(1450)^0$ component with the isobar model, implying a branching ratio ${\mathcal B}=(7.9\pm3.0)\times 10^{-7}$ for 
the quasi-two-body process $B^+\to\pi^+\rho(1450)^0\to\pi^+\pi^+\pi^-$ with the data 
${\mathcal B}=(1.52\pm0.14)\times 10^{-5}$~\cite{PDG-2018} for the three-body $B^+\to\pi^+\pi^+\pi^-$ decay.
With the same expressions for $B^+\to\pi^+\rho(1450)^0 \to\pi^+ K^+K^-$ but the subprocess $\rho(1450)^0\to\pi^+\pi^-$, 
we have the prediction 
\begin{eqnarray} 
 {\mathcal B} (B^\pm&\to&\pi^\pm\rho(1450)^0\to \pi^\pm \pi^+\pi^-)                               \nonumber \\
                                    &=&\big(  9.97\pm1.81(a^s_2+a^0_2+a^t_2) \pm0.98(\omega_B)  \nonumber\\  
                                    &\pm& 0.91(m^\pi_0+a^\pi_2)   \big)\times10^{-7},  
      \label{res-14502pipi}
\end{eqnarray}
which agree with the results $(7.9\pm3.0)\times 10^{-7}$ and $1.4^{+0.6}_{-0.9}\times10^{-6}$.

From the predictions for the quasi-two-body decays $B^\pm\to\pi^\pm\rho(1450)^0 \to\pi^\pm K^+K^-$ and 
$B^\pm\to\pi^\pm\rho(1450)^0\to \pi^\pm \pi^+\pi^-$ in this work, we have the ratio $R_{\rho(1450)}$ between their branching 
fractions as
\begin{eqnarray}
  R_{\rho(1450)} = \frac{{\mathcal B}(\rho(1450)^0\to K^+ K^-)}{{\mathcal B}(\rho(1450)^0 \to\pi^+\pi^-)} 
                       = 0.090\pm 0.017, \quad\;
   \label{PQCD-Rho-1450}
\end{eqnarray}
with the factorization relation 
$\Gamma(B^\pm\to \rho^0\pi^\pm\to h^+h^-\pi^\pm)\approx\Gamma(B^+\to \rho^0\pi^\pm){\mathcal B}(\rho^0\to h^+h^-)$, 
here $h=(\pi, K)$. The only error for $R_{\rho(1450)}$ comes from the uncertainty of $c^K_{\rho(1450)}$ because of the 
cancellation between the other errors of the two branching fractions in Eq.~(\ref{res-br-pi+1450}) and Eq.~(\ref{res-14502pipi}). 
That is to say, the increase or the decrease of the parameters that caused the errors will result in nearly identical change of the 
weight for the numerator and denominator of $R_{\rho(1450)}$.
The ratio $R_{\rho(1450)}$ can also be estimated from the coupling constants 
$g_{\rho(1450)^0\pi^+\pi^-}$ and $g_{\rho(1450)^0K^+K^-}$. 
With the relation $g_{\rho(1450)^0K^+K^-}\approx \frac12 g_{\rho(1450)^0\pi^+\pi^-}$~\cite{epjc39-41} one has
\begin{eqnarray}
  R_{\rho(1450)} &=& \frac{{\mathcal B}(\rho(1450)^0\to K^+ K^-)}{{\mathcal B}(\rho(1450)^0 \to\pi^+\pi^-)} \nonumber\\
                         &\approx& \frac{g^2_{\rho(1450)^0K^+K^-} (m^2_{\rho(1450)}-4m^2_K)^{3/2}}
                                                  {g^2_{\rho(1450)^0\pi^+\pi^-}(m^2_{\rho(1450)}-4m^2_\pi)^{3/2}} \nonumber\\
                        &=& 0.107.
   \label{th-Rho-1450}
\end{eqnarray}
This value is consistent with the result in Eq.~(\ref{PQCD-Rho-1450}). 

The ratio $R_{\rho(1450)}$ has been measured by BaBar Collaboration with the result 
\begin{eqnarray}
  R_{\rho(1450)} &=&  \frac{{\mathcal B}(\rho(1450)^0\to K^+ K^-)}{{\mathcal B}(\rho(1450)^0 \to\pi^+\pi^-)} \nonumber\\
                          &=&  0.307 \pm 0.084 ({\rm stat}) \pm 0.082 ({\rm sys}).
   \label{ex-Rho-1450}
\end{eqnarray}
in the Dalitz plot analyses of $J/\psi \to \pi^+ \pi^- \pi^0$ and $J/\psi \to K^+ K^- \pi^0$ decays in Ref.~\cite{prd95-072007}.
This result is quite large comparing with the values in Eqs.~(\ref{PQCD-Rho-1450})-(\ref{th-Rho-1450}). But one should note that 
the virtual contribution of $\rho(770)^0\to K^+ K^-$ has been ignored for the channel $J/\psi \to K^+ K^- \pi^0$ in~\cite{prd95-072007}. 
We argue that the large value for $R_{\rho(1450)}$ from BaBar could be understood as that the virtual contribution of $\rho(770)^0$ 
for $K^+ K^-$ has probably been taken into account for the process $\rho(1450)^0\to K^+ K^-$.

\begin{figure}[tbp]  
\centerline{\epsfxsize=8cm \epsffile{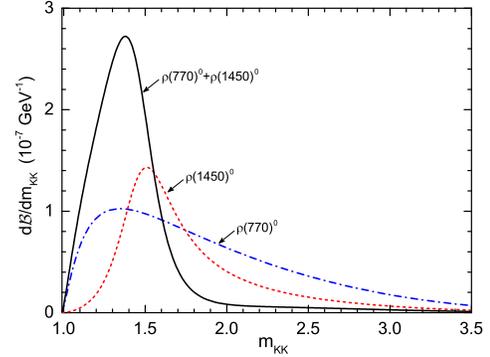}}
\vspace{-0.5cm}
\caption{The differential branching fractions for the quasi-two-body decays $B^\pm\to\pi^\pm\rho(770)^0\to \pi^\pm K^+K^-$,
              $B^\pm\to\pi^\pm\rho(1450)^0\to \pi^\pm K^+K^-$ and $B^\pm\to\pi^\pm(\rho(770)^0+\rho(1450)^0)\to \pi^\pm K^+K^-$.}
\label{fig-dep-br}
\end{figure}

The virtual contribution for $K^+K^-$ from the tail of the resonance $\rho(770)^0$, which has not been taken into the decay 
amplitudes of $B^\pm \to \pi^\pm K^+K^-$ and other charmless three-body $B$ meson decays involving kaon pair in  the 
experimental studies, is about $1.5$ times of the contribution from $\rho(1450)^0$ for the charged kaon pair 
as shown in Eq.~(\ref{res-br-pi+R770}) for the branching fractions. 
The predicted result $\left(1.31\pm0.27\right)\times10^{-7}$ is roughly $2.5\%$ of the total branching fraction of the corresponding 
three-body decay and larger than the fit fraction $(0.3\pm0.1\pm0.1)\%$ in the same channel for $\phi(1020)$ in~\cite{prl123-231802}, 
and should not be ignored 
in the studies of the three-body decays like $B^\pm \to \pi^\pm K^+K^-$. The large virtual contribution of $\rho(770)^0$ for the charged 
kaon pair is mainly due to its large decay width which result in a relative dispersive distribution of the differential branching fraction 
as shown in Fig.~\ref{fig-dep-br}, where the red short dash line is for the decay $B^\pm\to\pi^\pm\rho(1450)^0\to \pi^\pm K^+K^-$, 
the blue dash-dot line is for the process $B^\pm\to\pi^\pm\rho(770)^0\to \pi^\pm K^+K^-$ and the black solid line is for 
$B^\pm\to\pi^\pm(\rho(770)^0+\rho(1450)^0)\to \pi^\pm K^+K^-$.

In Fig.~\ref{fig-dep-br}, the peak of the differential branching fraction for $B^\pm\to\pi^\pm\rho(1450)^0\to \pi^\pm K^+K^-$ 
is around the pole mass of the $\rho(1450)$ resonance as expected. There is also a bump at about 
$1.35$~GeV of the invariant mass of kaon pair for the process $B^\pm\to\pi^\pm\rho(770)^0\to \pi^\pm K^+K^-$. 
This bump is generated by the the tail of the BW formula for 
$\rho(770)$ along with the phase space factors $\overrightarrow{q_K}$ and $\overrightarrow{q_\pi}$ in the 
Eqs.~(\ref{def-q-K})-(\ref{def-q-pi}). When we investigate the total contributions from $\rho(1450)^0$ plus $\rho(770)^0$, 
the bump from $\rho(770)^0$ will disappear and an enhanced peak around $1.4$ GeV is the only one existed as shown 
by the solid line in Fig.~\ref{fig-dep-br}. This makes us conclude that the absence of the virtual contribution from $\rho(770)^0$ for 
the decay amplitudes of $B^\pm\to\pi^\pm K^+K^-$ could probably enhance the experimental result for the 
$K^+K^-$ from the resonance $\rho(1450)^0$.

\begin{figure}[tbp]  
\centerline{\epsfxsize=8cm \epsffile{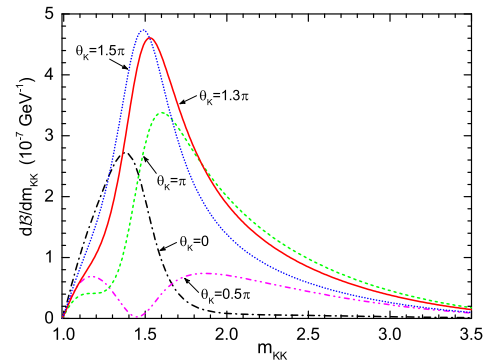}}
\vspace{-0.5cm}
\caption{The differential branching fractions for $B^\pm\to\pi^\pm(\rho(770)^0+\rho(1450)^0)\to \pi^\pm K^+K^-$ 
               with the phases $\theta_K=\{0, 0.5\pi,\pi,1.3\pi,1.5\pi\}$. }
\label{fig-dep-inter}
\end{figure}

In principle, the interference between $\rho(770)^0$ and $\rho(1450)^0$ can increase or decrease the total contributions from these 
two resonances. Then we adopt the same expressions and parameters of time-like form factor for the $\rho(770)^0$ and 
$\rho(1450)^0$ components in Eq.~(\ref{def-F-K+}), but add a phase $\theta_K$ between these two components, namely
\begin{eqnarray} 
  &~&F^{\rho(770+1450)}_K\nonumber\\
   &~&\quad = \frac12\left[c^K_{\rho(770)} {\rm GS}_{\rho(770)}+e^{{\rm i}\theta_K}c^K_{\rho(1450)} {\rm GS}_{\rho(1450)}\right]\!.\;\;
  \label{def-F-77+145}
\end{eqnarray}
When $\theta_K=1.3\pi$, we find the largest interferential branching fraction,
with ${\mathcal B}\approx 3.64\times10^{-7}$ for the $(\rho(770)^0+\rho(1450)^0)\to K^+K^-$ of the 
$B^\pm \to \pi^\pm K^+K^-$ decays. The differential branching fraction curves for 
$B^\pm\to\pi^\pm(\rho(770)^0+\rho(1450)^0)\to \pi^\pm K^+K^-$ with $\theta_K=\{0, 0.5\pi,\pi,1.3\pi,1.5\pi\}$ are 
shown in the Fig.~\ref{fig-dep-inter}, the red solid line is corresponding to $\theta_K=1.3\pi$.

The resonance $\omega(782)$, different from $\rho(770)$, has a narrow full width $\Gamma=8.49\pm0.08$ MeV.  In spite of the relation 
$g_{\omega(782)K^+K^-}=g_{\rho(770)^0K^+K^-}$~\cite{epjc39-41}, one can expect a small contribution from the tail of 
$\omega(782)$ for the $K^+K^-$ in the $B^\pm \to \pi^\pm K^+K^-$ decays.  As a result, we predict the branching fraction to be 
${\mathcal B}_v\approx 2.32 \times10^{-8}$ for the decay $B^+\to\pi^+\omega(782)\to \pi^+K^+K^-$, which is much smaller 
than the branching fractions for the decays $B^\pm \to \pi^\pm K^+K^-$ with the subprocesses $\rho(770)^0\to K^+K^-$ and 
$\rho(1450)^0\to K^+K^-$.  
The excited state $\rho(1700)^0$ has been studied in the quasi-two-body decays in PQCD approach with 
$\rho(1700)^0\to \pi^+\pi^-$~\cite{prd96-036014,prd96-093011}. With the Eq.~(\ref{th-Rho-1450}) and the relation 
$g_{\rho K^+K^-}\approx\frac12 g_{\rho\pi^+\pi^-}$ but the replacement $\rho(1450)^0\to \rho(1700)^0$, we estimate the 
ratio between the branching fractions of $\rho(1700)^0 \to K^+K^-$ and $\rho(1700)^0\to\pi^+\pi^-$ as 
\begin{eqnarray}
  R_{\rho(1700)} &=& \frac{{\mathcal B}(\rho(1700)^0\to K^+ K^-)}{{\mathcal B}(\rho(1700)^0 \to\pi^+\pi^-)} \nonumber\\
                         &\approx& \frac{g^2_{\rho(1700)^0K^+K^-} (m^2_{\rho(1700)}-4m^2_K)^{3/2}}
                                                  {g^2_{\rho(1700)^0\pi^+\pi^-}(m^2_{\rho(1700)}-4m^2_\pi)^{3/2}}  \nonumber\\
                        &=& 0.143,
   \label{th-Rho-1700}
\end{eqnarray}
with $m_{\rho(1700)}=1.72$~GeV~\cite{PDG-2018}. In consideration of the ${\mathcal B}=2.81^{+0.63}_{-0.66}\times10^{-7}$ 
in~\cite{prd96-036014} for the decay $B^+\to\pi^+\rho(1700)^0 \to\pi^+\pi^+\pi^-$, one could expect a branching fraction 
${\mathcal B}\approx4.02^{+0.90}_{-0.94}\times10^{-8}$ for the decay $B^+\to\pi^+\rho(1700)^0$ with the subprocess 
$\rho(1700)^0 \to K^+K^-$, which is close to half of the contribution from $\rho(1450)^0 \to K^+K^-$.
Then we have about $5\%$ of the total branching fraction, but still much less than the $(30.7\pm1.2\pm0.9)\%$ from 
LHCb~\cite{prl123-231802}, 
of the three-body decays $B^\pm \to \pi^\pm K^+K^-$ from the resonances $\rho(770)^0$, $\rho(1450)^0$ and $\rho(1700)^0$ 
when neglecting the interferences between them.

\section{Summary}
\label{sec-sum}
We analyzed the quasi-two-body decays $B^+\to\pi^+\rho(770,1450)^0 \to\pi^+ K^+K^-$ in the PQCD approach. 
Our results and analyses do not support that large branching fraction contributed by the resonance $\rho(1450)^0$ for the decays 
$B^\pm \to \pi^\pm K^+K^-$. The predictions of the branching fractions are ${\mathcal B}=(8.96\pm2.61)\times10^{-8}$ and 
${\mathcal B}_v=(1.31\pm0.27)\times10^{-7}$ for the quasi-two-body decays with the subprocesses $\rho(1450)^0 \to K^+K^-$ and 
$\rho(770)^0 \to K^+K^-$, respectively. 

The virtual contribution from the BW tail of $\rho(770)^0$ for $K^+K^-$, which has not been taken into the decay amplitudes of the 
charmless three-body $B$ meson decays involving kaon pair, was found about $1.5$ times of the contribution from the resonance 
$\rho(1450)^0$ in this work, and is roughly $2.5\%$ of the total branching fraction for the three-body decays 
$B^\pm \to \pi^\pm K^+K^-$.

There is a bump, which is generated by the tail of the BW formula and the phase space factors, at about $1.35$~GeV 
of the invariant mass of kaon pair for $B^\pm\to\pi^\pm\rho(770)^0\to \pi^\pm K^+K^-$. This bump will disappear 
and the enhancement round $1.4$ GeV will be the result when we investigated the total contributions from $\rho(1450)^0$ 
together with $\rho(770)^0$. This means that the absence of $\rho(770)^0\to K^+K^-$ in the decay amplitude of a three-body $B$ 
decays could probably result in a larger proportion for the resonance $\rho(1450)^0$ in experimental amplitude analysis.

\begin{acknowledgments}
This work was supported in part by the National Natural Science Foundation of China under Grants No.~11547038 and No.~11575110.
\end{acknowledgments}



\begin{thebibliography}{99}

\bibitem{npps199-341}
  K.~S.~F.~F. Guimar\~{a}es,  {\it et al.}, 
  Nucl.\ Phys.\ Proc.\ Suppl. {\bf 199}, 341 (2010).

\bibitem{prd84-094001}
  P.~C.~Magalh\~{a}es {\it et al.},
  Phys.\ Rev.\ D {\bf 84}, 094001 (2011).
  
\bibitem{1512-09284}
  I. Bediaga and P. C. Magalh\~aes,
  arXiv:1512.09284 [hep-ph].

\bibitem{prd89-094013} 
  I.~Bediaga, T.~Frederico and O.~Louren\c{c}o,
  Phys.\ Rev.\ D {\bf 89}, 094013 (2014).
  
\bibitem{pr135-B551} 
  G.~N.~Fleming,
  Phys.\ Rev.  {\bf 135}, B551 (1964).
  
\bibitem{pr166-1731} 
  D.~Morgan,
  Phys.\ Rev.  {\bf 166}, 1731 (1968).
  
\bibitem{prd11-3165} 
  D.~Herndon, P.~Soding and R.~J.~Cashmore, 
  Phys.\ Rev. {\bf D11}, 3165 (1975).

\bibitem{dalitz} 
  R.~H.~Dalitz,
  Phys.\ Rev.\  {\bf 94}, 1046 (1954).

\bibitem{prl123-231802} 
  R.~Aaij {\it et al.} (LHCb Collaboration),
  Phys.\ Rev.\ Lett.\  {\bf 123}, 231802 (2019).

\bibitem{PDG-2018} 
  M.~Tanabashi {\it et al.} (Particle Data Group),
  Phys.\ Rev.\ D {\bf 98}, 030001 (2018).

\bibitem{prd96-031101} 
  C.-L.~Hsu {\it et al.} (Belle Collaboration),
  Phys.\ Rev.\ D {\bf 96}, 031101(R) (2017).
 
\bibitem{prl99-221801} 
  B.~Aubert {\it et al.} (BaBar Collaboration),
  Phys.\ Rev.\ Lett.\  {\bf 99}, 221801 (2007).

\bibitem{prl124-031801} 
  R.~Aaij {\it et al.} (LHCb Collaboration),
  Phys.\ Rev.\ Lett.\  {\bf 124}, 031801 (2020).

\bibitem{prd101-012006}
  R.~Aaij {\it et al.} (LHCb Collaboration),
  Phys.\ Rev.\ D {\bf 101}, 012006 (2020).

\bibitem{prd76-094006} 
  H.~Y.~Cheng, C.~K.~Chua and A.~Soni,
  Phys.\ Rev.\ D {\bf 76}, 094006 (2007).

\bibitem{prd88-114014} 
  H.~Y.~Cheng and C.~K.~Chua,
  Phys.\ Rev.\ D {\bf 88}, 114014 (2013).

\bibitem{2003-03754} 
  Z.~T.~Zou, Y.~Li, Q.~X.~Li and X.~Liu,
  Eur.\ Phys.\ J.\ C {\bf 80}, 394 (2020).

\bibitem{epjc79-792} 
  Z.~Rui, Y.~Li and H.~Li,
  Eur.\ Phys.\ J.\ C {\bf 79}, 792 (2019).

\bibitem{prd96-113003} 
  D.~Boito,  {\it et al.,}  
  Phys.\ Rev.\ D {\bf 96}, 113003 (2017).
  
\bibitem{prd67-034012} 
  C.~K.~Chua, W.~S.~Hou, S.~Y.~Shiau and S.~Y.~Tsai,
  Phys.\ Rev.\ D {\bf 67}, 034012 (2003).
  
\bibitem{prd94-072001} 
  R.~Aaij {\it et al.} (LHCb Collaboration),
  Phys.\ Rev.\ D {\bf 94}, 072001 (2016).

\bibitem{plb791-342} 
  W.~F.~Wang and J.~Chai,
  Phys.\ Lett.\ B {\bf 791}, 342 (2019).

\bibitem{BW-model}
  G.~Breit and E.~Wigner,
  Phys.\ Rev.\  {\bf 49}, 519 (1936).
  
\bibitem{prd15-3196} 
  A.~J.~Pawlicki {\it et al.}, 
  Phys.\ Rev.\ D {\bf 15}, 3196 (1977).

\bibitem{prd22-2595}  
  D.~Cohen {\it et al.}, 
  Phys.\ Rev.\ D {\bf 22}, 2595 (1980).

\bibitem{plb669-217} 
  R.~R.~Akhmetshin {\it et al.} (CMD-2 Collaboration),
  Phys.\ Lett.\ B {\bf 669}, 217 (2008).
  
\bibitem{prd88-032013} 
  J.~P.~Lees {\it et al.} (BaBar Collaboration),
  Phys.\ Rev.\ D {\bf 88}, 032013 (2013).
  
\bibitem{prd94-112006}  
  M.~N.~Achasov {\it et al.},
  Phys.\ Rev.\ D {\bf 94}, 112006 (2016).

\bibitem{plb779-64} 
  E.~A.~Kozyrev {\it et al.},
  Phys.\ Lett.\ B {\bf 779}, 64 (2018).
  
\bibitem{prd99-032001} 
  M.~Ablikim {\it et al.} (BESIII Collaboration),
  Phys.\ Rev.\ D {\bf 99}, 032001 (2019).

\bibitem{epjc78-897} 
  J.~R.~Pel\'aez and A.~Rodas,
  Eur.\ Phys.\ J.\ C {\bf 78}, 897 (2018).
  
\bibitem{plb504-6}
  Y.~Y.~Keum, H.~n.~Li and A.~I.~Sanda,
  Phys.\ Lett. {\bf B504}, 6 (2001).

\bibitem{prd63-054008}
  Y.~Y.~Keum, H.~n.~Li and A.~I.~Sanda,
  Phys.\ Rev.\ D {\bf 63}, 054008 (2001).

\bibitem{prd63-074009}
  C.~D.~L\"u, K.~Ukai and M.~Z.~Yang,
  Phys.\ Rev.\ D {\bf 63}, 074009 (2001).

\bibitem{ppnp51-85}
  H.~n.~Li,
  Prog.\ Part.\ Nucl.\ Phys.\  {\bf 51}, 85 (2003).
  
\bibitem{prd75-074017} 
  X.~Liu, B.~Zhang, L.~L.~Shen and S.~L.~Zhu,
  Phys.\ Rev.\ D {\bf 75}, 074017 (2007).
 
\bibitem{plb763-29} 
  W.~F.~Wang and H.~n.~Li,
  Phys.\ Lett.\ B {\bf 763}, 29 (2016).

\bibitem{jhep2003-162} 
  W.~F.~Wang, J.~Chai and A.~J.~Ma, 
  J. High Energy Phys. 03 (2020) 162.
  
\bibitem{plb788-468} 
  W.~F.~Wang,
  Phys.\ Lett.\ B {\bf 788}, 468 (2019).
    
\bibitem{prd96-036014}  
  Y.~Li, A.~J.~Ma, W.~F.~Wang and Z.~J.~Xiao,
  Phys.\ Rev.\ D {\bf 96}, 036014 (2017).
  
\bibitem{prd95-056008} 
  Y.~Li, A.~J.~Ma, W.~F.~Wang and Z.~J.~Xiao,
  Phys.\ Rev.\ D {\bf 95}, 056008 (2017).

\bibitem{prd96-093011} 
  A.~J.~Ma, Y.~Li, W.~F.~Wang and Z.~J.~Xiao,
  Phys.\ Rev.\ D {\bf 96}, 093011 (2017).
  
\bibitem{npb923-54}
  A.~J.~Ma, Y.~Li, W.~F.~Wang and Z.~J.~Xiao,
  Nucl.\ Phys.\ B {\bf 923}, 54 (2017).
  
\bibitem{epjc39-41} 
  C.~Bruch, A.~Khodjamirian and J.~H.~K\"uhn,
  Eur.\ Phys.\ J.\ C {\bf 39}, 41 (2005).

\bibitem{npb250-517} 
  J.~Gasser and H.~Leutwyler,
  Nucl.\ Phys.\ B {\bf 250}, 517 (1985).

\bibitem{prl21-244} 
  G.~J.~Gounaris and J.~J.~Sakurai,
  Phys.\ Rev.\ Lett.\  {\bf 21}, 244 (1968).
  
\bibitem{prd86-032013} 
  J.~P.~Lees {\it et al.} (BaBar Collaboration),
  Phys.\ Rev.\ D {\bf 86}, 032013 (2012).

\bibitem{prd81-094014} 
  H.~Czy\.z, A.~Grzeli\'nska and J.~H.~K\"uhn,
  Phys.\ Rev.\ D {\bf 81}, 094014 (2010)
  
\bibitem{jetp129-386} 
  K.~I.~Beloborodov, V.~P.~Druzhinin and S.~I.~Serednyakov,
  J.\ Exp.\ Theor.\ Phys.\  {\bf 129}, 386 (2019).
  
\bibitem{prd76-074018} 
  A.~Ali {\it et al.}, 
  Phys.\ Rev.\ D {\bf 76}, 074018 (2007).
  
\bibitem{prd74-114009}
  B.~El-Bennich, A.~Furman, R.~Kami\'nski, L.~Le\'sniak and B.~Loiseau,
  Phys.\ Rev.\ D {\bf 74}, 114009 (2006).

\bibitem{prd79-094005}
  B.~El-Bennich {\it et al.}, 
  Phys.\ Rev.\ D {\bf 79}, 094005 (2009); Phys.\ Rev.\ D {\bf 83}, 039903(E) (2011).

\bibitem{Zemach} 
  C.~Zemach,
  Phys.\ Rev.\  {\bf 133}, B1201 (1964).
  
\bibitem{prd98-074512} 
  A.~Bazavov {\it et al.},
  Phys.\ Rev.\ D {\bf 98}, 074512 (2018).

\bibitem{prd78-114509} 
  C.~Allton {\it et al.} (RBC-UKQCD Collaboration),
  Phys.\ Rev.\ D {\bf 78}, 114509 (2008).
  
\bibitem{jhep1608-098} 
  A.~Bharucha, D.~M.~Straub and R.~Zwicky,
  JHEP {\bf 1608}, 098 (2016).

\bibitem{prd79-072006} 
  B.~Aubert {\it et al.} (BaBar Collaboration),
  Phys.\ Rev.\ D {\bf 79}, 072006 (2009).

\bibitem{prd95-072007} 
  J.~P.~Lees {\it et al.} (BaBar Collaboration),
  Phys.\ Rev.\ D {\bf 95}, 072007 (2017).

\end{thebibliography}
\end{document}